\documentclass[11pt]{article}
\usepackage{hyperref}
\usepackage{amssymb}
\usepackage{graphicx}
\usepackage{slashed}
\usepackage{amsmath}
\usepackage{breqn}
\usepackage{authblk}
\usepackage{dsfont}
\usepackage{cite}
\usepackage[section]{placeins}
\usepackage[a4paper, total={6.8in, 10in}]{geometry}
\numberwithin{equation}{section}
\newcommand{\be}{\begin{equation}}
\newcommand{\ee}{\end{equation}}
\newcommand{\bea}{\begin{eqnarray}}
\newcommand{\eea}{\end{eqnarray}}
\newcommand{\ba}{\begin{aligned}}
\newcommand{\ea}{\end{aligned}}
\begin{document}
\title{Regularization of central forces with damping  in two and three-dimensions}

\author{E. Harikumar \thanks{harisp.uoh@nic.in} and Suman Kumar Panja \thanks{sumanpanja19@gmail.com}}
\affil{School of Physics, University of Hyderabad, \\Central University P.O, Hyderabad-500046, Telangana, India}
\author{Partha Guha\thanks{ partha.guha@ku.ac.ae}}
\affil{ Department of Mathematics \\ Khalifa University of Science and Technology
P.O. Box 127788, Abu Dhabi, UAE}
\maketitle

\begin{abstract}
Regularization of damped motion under central forces in two and three-dimensions are investigated and equivalent, undamped systems are obtained. The dynamics of a particle moving in 
$\frac{1}{r}$ potential and subjected to a damping force is shown to be regularized a la Levi-Civita. We then generalize this regularization mapping to the case of damped motion in the 
potential $r^{-\frac{2N}{N+1}}$. Further equation of motion of a damped Kepler motion in 3-dimensions is mapped to an oscillator with inverted sextic potential and couplings, in 4-dimensions using Kustaanheimo-Stiefel regularization method. It is shown that the strength of the sextic potential is given by the damping co-efficient of the Kepler motion. Using homogeneous Hamiltonian formalism, we establish the mapping between the Hamiltonian of these two models.  Both in 2 and 3-dimensions, we show that the regularized equation is non-linear, in contrast to undamped cases. Mapping of a particle moving in a harmonic potential subjected to damping to an undamped system with shifted frequency is then derived using Bohlin-Sudman transformation.
\end{abstract}

\section{Introduction}

Dissipation being unavoidable in natural systems, is of  intrinsic interest. The investigation of such systems has a long history and continues to be an active area of research\cite{dekker, hasse,weiss}. Damped harmonic oscillator is one of the systems that has been studied vigorously as a prototype of dissipative systems. Various approaches such as (i) coupling the system to a heat bath, (ii) use of Bateman-Caldirola-Kanai(BCK)\cite{bate,caldi,kanai} model which uses a time dependent Lagrangian/Hamiltonian have been developed and adopted for studying different aspects of dissipative systems. Each of these methods though having unique advantages, has some unsatisfactory features\cite{green, edw,ray, gitman}.

BCK model has been shown to be plagued  by  difficulties in interpretation even at the classical level, as the time dependent BCK-Lagrangian/Hamiltonian leading to correct damped equation of motion is shown to describe a variable mass system rather than a truly damped oscillator\cite{green, ray}.  It has been argued that the equivalence between BCK model and damped harmonic oscillator is not valid globally( i.e., not for all times) and they are equivalent only for finite time scales\cite{gitman}.\\

 Generalization of action principle which allows to include non-conservative systems in its ambit was 
presented long ago\cite{herglotz} and this method can be used to study dissipative systems. Use of contact manifolds to study dissipative systems is a new upsurge of interests among mathematicians 
\cite{Bravetti,jfc-guha,leon}.
Regularization is considered to be a tool for converting singularities of a system of differential equations
into regular ones. This is carried out by changing the dependent and/or independent variables of a system
of (singular) differential equations so that the new system is free of singularities\cite{StiSch}.
A well developed theory of regularization in celestial mechanics might be attributed to Bohlin, Sundman and Levi-Civita\cite{bohlin, L-C}.
The notion of Levi-Civita regularization is  now  used  for  the  regularization  of the  binary  collisions  in  
the  planar (2-dimensional)  Kepler  problem\cite{L-C}.  Generalization of these approach  to 3-dimensions was obtained by Kustaanheimo and Stiefel\cite{K-S}. These regularization schemes also lead to
linear differential equations. There are other schemes of regularization such as those developed by Moser\cite{mos} and
Ligon-Schaaf\cite{LS}. Moser showed that the flow of the $n$-dimensional  Kepler problem on the surface of constant negative
energy is conjugate to the geodesic flow on the unit tangent bundle of $S^n$. The Ligon-Schaaf regularization procedure
tackles together all the surfaces of negative energies.

\bigskip

Kepler problem is the most studied system which has singularity. Keplerian orbit where the separation of particle from the central mass goes to zero is called a collision orbit. Detailed analysis of such orbits are necessary for artificial satellite missions, at the starting point as well as at the final landing point in moon/planet/exoplanet, the separation distance vanishes. Another situation that requires regularization is the low energy orbits where 
there is no collision, but the probe passes very close to moon/planet to achieve boost (known as slingshot effect) where the separation is negligibly small compared to other length scales involved\cite{koon}. Regularization also becomes handy in dealing with perturbations to Kepler problem due to effects such as presence of a third body, change in the shape of bodies from spherical one assumed.

Motion under the influence of central force in presence of friction arise in systems such as Rydberg atom and binary stars 
and has been of interest. Effect of friction on the shape of orbits has been 
studied\cite{Bham,cas,ace}. Kepler problem with drag term, which is linear in velocity and inversely proportional 
to the square of the separation distance was studied and analytical solutions 
were obtained\cite{dan, dmit, leach}. In these studies, existence of a conserved quantity which is related to 
angular momentum was exploited in deriving the solutions. Further, for a specific choice of
coefficient of the drag term, this equation was shown to be related to that of harmonic oscillator\cite{dmit}. 
Kepler equations augmented with drag term was used to model the orbits of artificial satellites landing on other planets/moon, 
for orbits of re-entry in to earth's atmosphere as well as for analyzing motion of low orbit  satellites.  
In this paper, we study regularization of collision orbits of particle moving under the influence of inverse power 
law potentials and also subjected to velocity dependent damping, in 2 and 3 dimensions.

\bigskip

In this paper, we use the approaches of Levi-Civita and  Kustaanheimo and Stiefel to regularize problem of a particle moving in central potentials and subjected to damping in 2 and 3-dimensions.
In particular, we investigate the construction of equivalent model corresponding to systems with damping, viz: ~(i) particle in 2 and 3-dimensions, moving under the action of the  $\frac{1}{r}$ potential 
subjected to damping, and
(ii) particle in 2-dimensions moving under the influence of  
a potential of the form $r^{-2N/N+1}$ where $N\in {\mathds Z}$,  subjected to damping. 
We then study regularization of collision orbits of the particle in these situations.  
The equations of motion describing the damped motion, as well as the corresponding Lagrangian/Hamiltonian 
in the above  cases have explicit time-dependence. Thus for these systems, 
energy is not a conserved quantity and this hinders the direct application of 
Levi-Civita/K-S-transformations to these systems with damping. 
Also it is of interest to see how these equations transform under the 
re-parametrization of time, which is an integral part of these regularizations. 
The re-parametrization of time which makes the motion near the collision 
point ``slow" do affect the form of the equations of motion in terms of the new co-ordinates, 
through velocity and acceleration. Since in the case of damped systems we study, 
not only Lagrangian and Hamiltonian, but the 
equations of motion also dependent explicitly on the time,it is of interest to 
see how this explicit time dependence affect the regularization. In our analysis, 
we start with a Lagrangian (which can be related to a
BCK type, time dependent Lagrangian) whose equation of motion describes a particle 
subjected to Kepler potential, in addition to a velocity dependent damping.  
We first  map these equations  using a 
time dependent  point transformation such that the transformed equations 
follow as Euler-Lagrange equations from a time independent 
Lagrangian. This allows us to construct conserved energy, which in turn allow the implementation of the mapping of dynamics on a constant energy surface. We apply Levi-Civita map /K-S transformation 
to these equations, after re-expressing them in terms of complex co-ordinates/quaternions.  The equation in terms of the new complex co-ordinates/quaternions and time parameter is shown to describe a harmonic 
oscillator augmented by an inverted sextic potential in 2-dimensions/4-dimensions. \\

\smallskip

The regularization scheme known as Levi-Civita transformation for 2-dimensional Kepler problem and its generalization to 3-dimensions known as Kustaanheimo-Stiefel (K-S) transformation, do linearize the
differential equations describing the motion of a particle under the influence of gravitational force excreted by a central body and also removes the singularity of the equations when the separation of these two bodies vanishes. The procedure consists of {\it three steps}; 
\begin{enumerate}
\item In the {\it first step}, one implements a re-parametrization of time so that  the velocity defined in terms of  ``new" time variable, do not diverge as the separation distance approaches zero. One now uses chain rule and re-express the time derivatives of the position co-ordinates appearing in the differential equation in terms of the ``new" time variable. 
\item In the {\it second step}, one applies Levi-Civita conformal squaring (of the 
complex co-ordinates) for 2-dimensional case and K-S transformation which relates co-ordinates  of 3-dimensional space to square of quaternions. Thus the equations of motion are now re-expressed in terms of new co-ordinates and their derivatives with respect to ``new" time parameter. Thus obtained equation is regular but non-linear. 
\item In the {\it third or final step}, one starts with the conserved energy associated 
with the initial system, written in terms of kinetic and potential energies. One 
re-expresses the velocities appearing in kinetic part in terms of the derivative of new co-ordinates with respect to ``new" time parameter and co-ordinates in potential energy are re-written in terms of new co-ordinates. This conserved quantity is then used to re-express the equations of motion obtained in the second step above. Straight forward calculations then converts the equations of motion to be regular and linear one. This equation is then solved and by fixing the numerical value of the conserved quantity, one maps orbits of fixed energy to regular solutions of the linear equation.
\end{enumerate}

\bigskip

In this paper, we apply  these regularization methods to Kepler problem in 2 and 3 dimensions as well as to a generic inverse power law potential, all subjected to velocity dependent damping. We show that the mapping  leads to regular, but coupled equations. Further, the coupling terms all have the damping parameter as co-efficient.  We also study  the application of Bohlin-Sudman transformation to the equation describing  damped harmonic oscillator in 2-dimensions.  After applying a time dependent co-ordinate transformation, we map the damped harmonic motion to  that of a shifted harmonic oscillator.  As earlier, 
this allows us to define conserved energy which facilitate the mapping of dynamics from a constant energy surface. This equation, after re-expressing in terms of complex co-ordinates, are mapped to that of 2-dim Kepler problem by Bohlin-Sudman transformation\cite{bohlin}.\\

This paper is organized as follows. In the next section,  we  show that the Levi-Civita transformation maps the Kepler's equations  with damping in 2-dimensions to the equations of motion of a harmonic oscillator with an additional, inverted, sextic potential. This is done by first mapping the damped Kepler equations to an equivalent  set of equations which do not have explicit time dependence. These equations comes from a time independent Lagrangian and thus the corresponding energy is a conserved quantity. These equations are mapped by Levi-Civita regularization map to equations describing a harmonic oscillator, augmented with inverted, sextic potential.
In section 3, we use Levi-Civita map to obtain a regularized equation corresponding to equation describing a particle moving, in presence of friction,  under the generic inverse power law potential of the form $r^{-2N/N+1}$, $0\le 2N/N+1<2$ . In obtaining this mapping we follow the same steps as we have used in section 2, in deriving equivalence between 2-dimensional Kepler motion with drag to that of harmonic oscillator in presence of inverted  sextic potential. In both these cases, we express the conformal squaring of the co-ordinates (discussed above) as a matrix equation and further show that this equation can be expressed using a generic matrix connecting old and new co-ordinate. This matrix is written in terms of two permutation operators which do commute among themselves. In section 4, we study the Kepler problem in 3-dimensions in presence of damping. After a brief summary of quaternions, in subsection 4.2, we show that the K-S transformation maps the equations of motion of 3-dimensional Kepler problem with damping to that of a 4-dimensional harmonic oscillator with the inverted sextic potential. 
In subsection 4.3, we show the mapping of these two systems at the level of Hamiltonians, using homogeneous Hamiltonian formalism. Our concluding remarks are given in section 5.  In appendix A, we show the maping of damped harmonic motion to that of Kepler problem, using Bohlin-Sudman transformation.

\section{ Kepler problem in presence of damping in 2-dimensions: Levi-Civita map}

In this section, we apply Levi-Civita map to damped Kepler problem in 2-dimensions and obtain an equivalent but regularized formulation. We begin with a BCK type, time dependent Lagrangian whose equation of motion describes a particle moving in  Kepler potential and also subjected to  velocity dependent damping. We start by mapping these equations using a time dependent point transformation, so that the modified equations are Euler-Lagrange equations from a time independent Lagrangian. This permits us to build conserved energy, which allows us to apply the mapping of dynamics on a constant energy surface. After re-expressing these equations in terms of complex co-ordinates, we use the Levi-Civita map to get equations for a harmonic oscillator with inverted sextic potential and interactions. 

We note that the equation of motion following from a BCK type Lagrangian 
\be
L=e^{\lambda t}\left [\frac{m}{2}( {\dot x_{1}}^2+{\dot x_{2}}^2)+\frac{ke^{-\frac{3\lambda t}{2}}} {\tilde{r}}\right]
\ee
 where $\tilde{r}=\sqrt{x_{1}^2+x_{2}^2}$, $m=\frac{m_1+m_2}{m_1 m_2}$ and $k=Gm_1m_2$ is
\be
m{\ddot x}_i+\lambda m{\dot x}_i+\frac{ke^{-\frac{3\lambda t}{2}}} {\tilde{r}^3} x_i=0,~i=1,2,\label{dkp}
\ee
where ${\dot x}_i=\frac{dx_i}{dt}$ and ${\ddot x}_i=\frac{d^2x_i}{dt^2}$. These equations describe motion of a particle of mass $m_1$  
under the influence of a  gravitational potential of another body of mass $m_2$ as well as a damping force, in 
2-dimensions. Here when $\lambda\longrightarrow 0$, above Eqn.\ref{dkp} becomes equation of motion for well known Kepler problem in 2-dimensions. Note that this equation has explicit time dependence and the 
re-parametrization of time applied in the first step of regularization, will not remove this explicit dependence on the ``old" time.  Also, the energy of this system is not conserved and thus implementing the third step of 
regularization is not possible. In order to avoid these complications, we now re-write these equation of motion in terms of a new set of 
co-ordinates $(X_{1},X_{2})$ which are related to old co-ordinates through time dependent transformation given by 
\bea
X_1=x_1e^{\frac{\lambda t}{2}},\label{xX}\nonumber\\
X_2= x_2e^{\frac{\lambda t}{2}}.\label{yY}\label{mapeqn}
\eea
In terms of these co-ordinates, Eqn.(\ref{dkp}) become
\bea
m{\ddot X_{1}}-\frac{m\lambda^2}{4}X_1+\frac{kX_1}{(X_{1}^2+X_{2}^2)^{\frac{3}{2}}}=0,\label{Xeqn}\\
m{\ddot X_{2}}-\frac{m\lambda^2}{4}X_2+\frac{kX_2}{(X_{1}^2+X_{2}^2)^{\frac{3}{2}}}=0\label{Yeqn}.
\eea
These equations do not have explicit time dependence and first difficulty in implementing the regularization is now circumvented. 

It is easy to see that these are Euler-Lagrange equations following from the Lagrangian
\be
L=\frac{m}{2}({\dot X_1}^2+{\dot X_2}^2)+\frac{m\lambda^2}{8}(X_{1}^2+X_{2}^2)-\frac{m\lambda}{2}(X_1{\dot X_{1}}+X_2{\dot X_{2}})
+\frac{k}{r}\label{lag1}
\ee
Note that the third term can be re-written as $-\frac{m\lambda}{4}\frac{d}{dt}(X_{1}^2+X_{2}^2)$, a total derivative and thus will not contribute to equations of motion.  Following the standard procedure, we calculate the 
canonical conjugate momenta corresponding to $X_1$ and $X_2$ as
\bea
P_{X_{1}}=m{\dot X_{1}}-\frac{\lambda}{2}mX_1,\\
P_{X_{2}}=m{\dot X_{2}}-\frac{\lambda}{2}mX_2
\eea
and using these derive the Hamiltonian as
\be
{\cal H}=\frac{(P_{X_{1}}^2+P_{X_{2}}^2) }{2m}+\frac{\lambda}{2} (X_1P_{X_{1}}+X_2P_{X_{2}})-\frac{k}{r},\label{kham}
\ee
where we now use $r=\sqrt{X_{1}^2+X_{2}^2}$. This Hamiltonian does not have any explicit time dependence
and is invariant under time translations and hence total energy of this system is conserved. Thus, the second difficulty mentioned above in implementing regularization is also avoided.

We note that the  angular momentum $L=X_1P_{X_{2}}-X_2P_{X_{1}}$ is also conserved, as in the usual 
Kepler problem. For later purposes, we express (negative of) this Hamiltonian, which is a conserved quantity,
in terms of velocities and co-ordinates, viz:
\be
-{\cal E}=\frac{m}{2}({\dot X_{1}}^2+{\dot X_{2}}^2)-\frac{\lambda^2}{8}m(X_{1}^2+X_{2}^2)-\frac{k}{r}\label{h} .
\ee

We re-express the Eqn.(\ref{Xeqn}) and Eqn.(\ref{Yeqn}) in terms of complex variable $Z$ given by $Z=X_1+iX_2$ as
\be
m{\ddot Z}+\frac{k}{|Z|^3}Z=\frac{m\lambda^2}{4}Z,\label{Zeqn}
\ee
where $|Z|=\sqrt{X_{1}^2+X_{2}^2}=r$. Note that the derivative with respect to the time variable $t$ is represented by 
`overdot', in the above equations.

We now apply a re-parametrization of time and demand 
\footnote{Conserved angular momentum in damped Kepler problem and perturbed harmonic oscillator 
(see Eqn.(\ref{lag1}),Eqn.(\ref{secho})) allows us to equate these numbers and this leads 
to the relation in Eqn.(\ref{repara1}).}
\be
\frac{d}{dt}=\frac{c}{r}\frac{d}{d\tau}\label{repara1}.
\ee
Note here that the $c$ appearing in the above equation is a proportionality constant(and not the velocity of light).
Applying this re-parametrization, we find
\bea
\frac{dZ}{dt}&=&\frac{c}{r}\frac {dZ}{d\tau}\\
\frac{d^2Z}{dt^2}&=&\frac{c^2}{r^2}\frac{d^2Z}{d\tau^2}-\frac{c^2}{r^3}\frac{d r}{d\tau}\frac{dZ}{d\tau}.\label{derivative}
\eea
Using these, we re-express Eqn.(\ref{Zeqn}) as
\be
c^2(r\frac{d^2Z}{d\tau^2}-\frac{dr}{d\tau}\frac{dZ}{d\tau})+\mu Z=\frac{\lambda^2}{4}r^3Z,\label{taueqn}
\ee
where $\mu=k/m$.

Next we apply a co-ordinate transformation(Levi-Civita regularization) and re-write above equation in terms new complex 
co-ordinate $U=U_1+iU_2$ which is related to $Z$ as 
\be
Z=\gamma U^2\label{l-c}
\ee
which we can rewrite in matrix representation as following,
\bea
\left(\begin{array}{c}

X_1\\
X_2

\end{array}\right) =\gamma A(U)=\gamma \hat{U}_1
\left(\begin{array}{c}

U_1\\
U_2

\end{array}\right), \label{AU1}
\eea 
where 
\bea
A(U)=\hat{U}_1=\left(\begin{array}{cc}
U_1 & -U_2\\
U_2 & U_1
\end{array}\right).\label{AU2}
\eea
Which we can rewrite as follows,
\be
A(U)=\left(\mathcal{P}^{(1)}\textbf{U},\mathcal{P}^{(2)}\textbf{U} \right), \label{AU3}
\ee 
where we have considered following set of permutations,
\be
\mathcal{P}^{(1)}\textbf{U}=(U_1,U_2)^{T}, ~~~ \mathcal{P}^{(2)}\textbf{U} = (-U_2,U_1)^{T}. \label{Per1}
 \ee
 Which  have properties like,
 \be
 \mathcal{P}^{(i)}\mathcal{P}^{(j)} =  \mathcal{P}^{(j)}\mathcal{P}^{(i)} ~ and~~ (\mathcal{P}^{i})^{T}\mathcal{P}^{j}=\delta_{ij}. \label{Per2}
 \ee
In Eqn.(\ref{l-c}), $\gamma$ is constant having dimensions of inverse length(without lose of generality, we set this $\gamma$ 
to be one from now onwards). This sets $r=|Z|=U{\bar U}=\left|U\right|^2, r^2={\bar Z}Z$. We also find $\frac{dZ}{d\tau}=2U\frac{dU}{d\tau},$
~$\frac{d^2Z}{d\tau^2}=2(\frac{dU}{d\tau})^2+2U\frac{d^2U}{d\tau^2}$ and $\frac{dr}{d\tau}={\bar U}\frac{dU}{d\tau}+\frac{d{\bar U}}{d\tau} U$, using which, the Eqn.(\ref{taueqn}) becomes
\be
2c^2U{\bar U}\left(U\frac{d^2U}{d\tau^2}+\left(\frac{dU}{d\tau}\right)^2\right)-c^2\left({\bar U}\frac{dU}{d\tau}+\frac{d{\bar U}}{d\tau}U\right)2U\frac{dU}{d\tau}+\mu U^2=\frac{\lambda^2}{4}U^2({\bar U}U)^3.\label{222}
\ee 
This equation, after straight forward algebra gives
\be
2r\frac{d^2U}{d\tau^2}-2\left|\frac{dU}{d\tau}\right|^2U-\frac{\lambda^2}{4c^2} r^3U+\frac{\mu}{c^2} U=0.\label{u-eqn}
\ee
Next,we express the conserved quantity $-{\cal E}$ in terms of $\frac{d U}{d\tau}$. For this, we first re-express $-{\cal E}$ in terms of $Z$ and $\frac{dZ}{dt}$ as
 \be
 -{\cal E}=\frac{m}{2}\frac{d{\bar Z}}{dt}\frac{dZ}{dt}-\frac{m\lambda^2}{8}{\bar Z}Z-\frac{k}{|Z|}\label{com1}
 \ee
 and apply re-parametrization to replace derivative with respect to $t$ with derivative with respect to 
 $\tau$ and finally, re-express these terms using the derivative of $U$. This gives
 \bea
 -{\cal E}&=&\frac{m}{2}\frac{c^2}{r^2}\frac{d{\bar Z}}{d\tau}\frac{d Z}{d\tau}-\frac{m\lambda^2}{8}r^2-\frac{k}{r}\\
 &\equiv&\frac{2mc^2}{r}\left|\frac{dU}{d\tau}\right|^2-\frac{m\lambda^2}{8} r^2-\frac{k}{r}.
 \eea
  Using this, we re-write  $\left|\frac{dU}{d\tau}\right|^2$ in Eqn.(\ref{u-eqn}) and find
  \be
  \frac{d^2U}{d\tau^2}+\left(\frac{{\cal E}}{2mc^2}-\frac{3\lambda^2}{16c^2}r^2\right)U=0\label{sechoeqn} ,
  \ee
  where $r^2=\left|U\right|^4$. This equation describes a perturbed harmonic oscillator for $\lambda<<1$. To see this
  clearly, we re-write the above equation in $U_1$ and $U_2$ as
  \bea
  U_{1}^{\prime\prime}+\frac{1}{2c^2}\left(\frac{{\cal E}}{m}-\frac{3}{8}\lambda^2(U_{1}^2+U_{2}^2)^2\right)U_1=0\label{secho1}\\
  U_{2}^{\prime\prime}+\frac{1}{2c^2}\left(\frac{{\cal E}}{m}-\frac{3}{8}\lambda^2(U_{1}^2+U_{2}^2)^2\right)U_2=0\label{secho2}
  \eea
  where a `prime' over $U$ stands for derivative with respect to the new time variable, $\tau$. It is easy to see
  that these equations follow from the Lagrangian
  \be
  {\cal L}=\frac{m}{2}(U_{1}^{\prime 2}+U_{2}^{\prime 2})-\frac{{\cal E}}{4c^2}(U_{1}^2+U_{2}^2)
  +\frac{1}{32}\frac{m\lambda^2}{c^2}\left(U_{1}^2+U_{2}^2\right)^3.\label{secho}
  \ee
  We now see that the system described by the above Lagrangian is oscillator in 2-dimensions with inverted sextic potential(i.e., 
  with coefficients of $U_{1}^2$ and $U_{2}^2$ are negative) and couplings. Note that, for small $\lambda$, we can treat this 
  system as uncoupled harmonic oscillators with perturbations involving couplings and inverted sextic potential
  \footnote{Inserting $\gamma$ back into above equation, we find the coefficient of the last term in the 
  Lagrangian is $\frac{m\lambda^2\gamma^2}{32c^2}$.}.
  
  The Hamiltonian following from the above Lagrangian is given by
  \be
  H=\frac{P_{U_{1}}^2}{2m}+\frac{P_{U_{1}}^2}{2m}+\frac{{\cal E}}{4c^2}(U_{1}^2+U_{2}^2)
-\frac{1}{32}\frac{m\lambda^2}{c^2}\left(U_{1}^2+U_{2}^2\right)^3.\label{secho3}  
  \ee
Note that the $\lambda$ dependent term is with negative coefficient.

Thus we have shown that the equations of motion following from the Lagrangian in Eqn.(\ref{lag1}) are mapped by 
Levi-Civita map to equations following from the Lagrangian in Eqn. (\ref{secho}). This generalizes the equivalence of the Kepler problem to the harmonic oscillator in 2-dimensions to the equivalence of the  Kepler problem 
in presence of a damping force, to a perturbed harmonic oscillator. This is shown by first mapping the  Eqn.(\ref{dkp}) describing the damped Kepler equations in 2-dimensions to the Eqns.(\ref{Xeqn},\ref{Yeqn})
(which follow from the Lagrangian given in Eqn.(\ref{lag1})), which are then mapped to that of harmonic oscillator in 2-dimensions with specific couplings and inverted sextic potential. This mapping is obtained for the 
surface defined by the constant value of ${\cal E}$. From Eqn.(\ref{com1}), we note that this conserved quantity reduces to energy of the Kepler system in the limit $\lambda\to 0$ and thus, in the limit $\lambda\to 0$, we get back the well known equivalence between Kepler motion and harmonic oscillations in 2-dimensions, on the constant energy surface.

\section{ Mapping of motion in  $\frac{1}{r^{\frac{2N}{{N+1}}}}, 0\le 2N/N+1<2$ potential with damping  in 2-dimensions}

In this section, we show that the Levi-Civita mapping  regularizes the  damped motion in generic inverse power law potential given by   $\frac{1}{r^{\frac{2N}{{N+1}}}}, 0\le 2N/N+1<2$. In the undamped case, such a generalization was shown in\cite{budd}. After expressing the transformations between old and new co-ordinates as a matrix equation, in\cite{budd}, this matrix was written using a pair of permutation operators. The commuting nature of these permutation operators was crucial in proving various identities which in turn were used for this generalization of Levi-Civita transformation. Here, we apply this generalization to the system moving under the influence of the potential $\frac{1}{r^{\frac{2N}{{N+1}}}}, 0\le 2N/N+1<2$ and subjected to damping.

We start from a BCK type Lagrangian given by 
\be
L=e^{\lambda t}\left [\frac{m}{2}( {\dot x_{1}}^2+{\dot x_{2}}^2)+\frac{ke^{-\frac{2N+1}{N+1}\lambda t}} {\tilde{r}^{\frac{2N}{N+1}}}\right]
\ee
 where $\tilde{r}=\sqrt{x_{1}^2+x_{2}^2}$, $m=\frac{m_1+m_2}{m_1 m_2}$ and $k=Gm_1m_2$ . The equations of motion following from this Lagrangian are
\be
m{\ddot x}_i+\lambda m{\dot x}_i+\left(\frac{2N}{N+1}\right) \frac{ke^{-\frac{2N+1}{N+1}\lambda t}} {\tilde{r}^{\frac{4N+2}{N+1}}} x_i=0,~i=1,2,\label{gp}
\ee
where ${\dot x}_i=\frac{dx_i}{dt}$ and ${\ddot x}_i=\frac{d^2x_i}{dt^2}$. Here when $N=1$, we get back equation of motions for perturbed Kepler problem given in Eqn.\ref{dkp} and additionally when $\lambda\longrightarrow 0$ , above Eqn.\ref{gp} becomes equation of motion for well known Kepler problem in 2-dimensions

We now re-write these equation of motion in terms a new set of 
co-ordinates $(X_{1},X_{2})$ which are related to old co-ordinates through time dependent transformation given by 
\bea
X_1=x_1e^{\frac{\lambda t}{2}},\label{gp1}\nonumber\\
X_2= x_2e^{\frac{\lambda t}{2}}.\label{gp2}\label{mapeqn}
\eea
In terms of these co-ordinates, Eqn.(\ref{gp}) become
\be
m{\ddot X_{i}}-\frac{m\lambda^2}{4}X_i+\left(\frac{2N}{N+1}\right) \frac{kX_i}{r^{\frac{4N+2}{N+1}}}=0,~ i=1,2 \label{gp3}
\ee
where $r=\sqrt{X_{1}^2+X_{2}^2}$.
Here we can see that these equations following from the Lagrangian
\be
L=\frac{m}{2}({\dot X_1}^2+{\dot X_2}^2)+\frac{m\lambda^2}{8}(X_{1}^2+X_{2}^2)-\frac{m\lambda}{2}(X_1{\dot X_{1}}+X_2{\dot X_{2}})
+\frac{k}{r^{\frac{2N}{N+1}}},\label{gp4}
\ee
which has no explicit time dependence. Note that here $N=1$ leads to Lagrangian of perturbed Kepler problem given in Eqn.(\ref{lag1}) and third term as a total derivative term will not contribute to equation of motions.  Following the standard procedure, we calculate the 
canonical conjugate momenta corresponding to $X_1$ and $X_2$ as
\bea
P_{X_{1}}=m{\dot X_{1}}-\frac{\lambda}{2}mX_1\\
P_{X_{2}}=m{\dot X_{2}}-\frac{\lambda}{2}mX_2
\eea
and using these we derive the Hamiltonian as
\be
{\cal H}=\frac{(P_{X_{1}}^2+P_{X_{2}}^2) }{2m}+\frac{\lambda}{2} (X_1P_{X_{1}}+X_2P_{X_{2}})-\frac{k}{r^{\frac{2N}{N+1}}},\label{gp5}
\ee
where  $r=\sqrt{X_{1}^2+X_{2}^2}$ and at $N=1$ above Hamiltonian leads to Eqn.(\ref{kham}). This Hamiltonian does not have any explicit time dependence
and is invariant under time translations and hence total energy of this system is conserved. For later purposes, we express (negative of) this Hamiltonian, which is a conserved quantity,
in terms of velocities and co-ordinates, viz:
\be
-{\cal E}=-\frac{E}{m}=\frac{1}{2}({\dot X_{1}}^2+{\dot X_{2}}^2)-\frac{\lambda^2}{8}(X_{1}^2+X_{2}^2)-\frac{\mu}{r^{\frac{2N}{N+1}}}\label{gp6},
\ee
where $\frac{k}{m}=\mu$

We re-express the Eqn.(\ref{gp3}) in terms of complex variable $Z$ given by $Z=X_1+iX_2$ as
\be
m{\ddot Z}+\left(\frac{2N}{N+1}\right)\frac{k}{|Z|^{\frac{4N+2}{N+1}}}Z=\frac{m\lambda^2}{4}Z,\label{gp7}
\ee
where $|Z|=\sqrt{X_{1}^2+X_{2}^2}=r$. Note that the derivative with respect to the time variable $t$ is represented by 
`overdot', in the above equations.

We now apply a re-parametrization of time to avoid presence of singularity from the equations of motion  and demand 

\be
\frac{d}{dt}=\frac{1}{g(r)}\frac{d}{d\tau}, ~~ where~ g(r)=(N+1)^2 r^{\frac{2N}{N+1}} \label{gp8}
\ee
Note here that the $N=1$ leads to Eqn.(\ref{repara1}), time re-parametrization used for perturbed Kepler problem.
Applying this re-parametrization, we find
\bea
\frac{dZ}{dt}&=&\frac{1}{(N+1)^2 r^{\frac{2N}{N+1}}}\frac {dZ}{d\tau}\\
\frac{d^2Z}{dt^2}&=&\frac{1}{(N+1)^4\left(r^{\frac{2N}{N+1}}\right)^2}\frac{d^2Z}{d\tau^2}-\left(\frac{2N}{(N+1)^5}\right) \frac{1}{r^{\frac{5N+1}{N+1}}}\frac{d r}{d\tau}\frac{dZ}{d\tau}.\label{gp9}
\eea
Using these, we re-express Eqn.(\ref{gp7}) as
\be
\frac{1}{(N+1)^4 r^{\frac{4N}{N+1}}}\frac{d^2Z}{d\tau^2}-\frac{2N}{(N+1)^5}\frac{1}{r^{\frac{5N+1}{N+1}}}\frac{dr}{d\tau}\frac{dZ}{d\tau}+\left( \frac{2N}{N+1} \right) \frac{\mu}{r^{\frac{4N+2}{N+1}}} Z=\frac{\lambda^2}{4}Z,\label{gp10}
\ee
where $\mu=k/m$.

Next we apply a co-ordinate transformation(Levi-Civita regularization) and consider following relations (\cite{budd}),
\be
\textbf{Z}=\hat{U}_N\textbf{U} \label{gplc}
\ee
where,
\be
\textbf{U}^{(N)}=\hat{U}_{N-1}\textbf{U} ~~ and~~ \hat{U}_N=\left(\mathcal{P}^{(1)}\textbf{U}^{(N)},\mathcal{P}^{(2)}\textbf{U}^{(N)} \right), ~N \geq 1 \label{gplc1}
\ee 
with $\hat{U}_0= I$ and $\hat{U}_1=A(U)$ as explained in Eqn.(\ref{AU2}). Here we consider
same definition of permutation matrices as discussed previously in Eqn.(\ref{Per1}) i.e,
\be
\mathcal{P}^{(1)}\textbf{U}=(U_1,U_2)^{T}, ~~~ \mathcal{P}^{(2)}\textbf{U} = (-U_2,U_1)^{T}, \label{gpPer1}
 \ee
 which  have following properties like,
 \be
 \mathcal{P}^{(i)}\mathcal{P}^{(j)} =  \mathcal{P}^{(j)}\mathcal{P}^{(i)} ~ and~~ (\mathcal{P}^{i})^{T}\mathcal{P}^{j}=\delta_{ij} \label{gpPer2}.
 \ee
We also consider following relations (\cite{budd}),
\be
\hat{U}^{T}_{N} \hat{U}_N=\hat{U}_N\hat{U}^{T}_{N}=R^{2N}I_{2\times2} ~ and~ (\textbf{U}^{(N+1)})^{\prime}=(N+1)\hat{U}_N\textbf{U}^{\prime} \label{gpPer3},
\ee
where $(\prime)$ denotes time derivative with respect to fictious time $\tau$.
 
 These set $r=R^{N+1}, r^{\prime}=(N+1)R^{N}R^{\prime}$. We also find (in matrix representation) $\textbf{Z}^{\prime}=(N+1)\hat{U}_N \textbf{U}^{\prime}$,
~$\textbf{Z}^{\prime \prime}=(N+1)\hat{U}^{\prime}_N \textbf{U}^{\prime}+(N+1)\hat{U}_N \textbf{U}^{\prime \prime}$, using which, we can rewrite Eqn.(\ref{gp10}) in matrix representation as,
\be
(N+1)\hat{U}^{\prime}_N \textbf{U}^{\prime}+(N+1)\hat{U}_N\textbf{U}^{\prime \prime} -\frac{1}{R}(2N)(N+1)R^{\prime}(\hat{U}_N \textbf{U}^{\prime}) -\frac{\lambda^2}{4}(N+1)^4 R^{4N}\hat{U}_N \textbf{U}+\frac{\mu(2N)(N+1)^3}{R^2}\hat{U}_N \textbf{U}=0 \label{gpPer4}
\ee 
This equation, after straight forward algebra gives
\be
\textbf{U}^{\prime \prime}-\frac{\lambda^2}{4}(N+1)^{3}R^{4N}\textbf{U}+\left\lbrace \frac{\mu(2N)(N+1)^2}{R^2}-\frac{N(\sum (U^{\prime}_i)^2)}{R^2}\right\rbrace \textbf{U}=0. \label{gpPer5}
\ee
Here for $N=1$, we get back Eqn.(\ref{222}) with c being $\frac{1}{4}$.
Next,we express the conserved quantity $-{\cal E}$ in terms of new-coordinates. For this, we first re-express $-{\cal E}$ in terms of $Z$ and $\frac{dZ}{dt}$ as
 \be
 -{\cal E}=\frac{1}{2}\frac{d{\bar Z}}{dt}\frac{dZ}{dt}-\frac{\lambda^2}{8}{Z}{\bar Z}-\frac{\mu}{|Z|^{\frac{2N}{N+1}}}\label{gpce}
 \ee
 and apply re-parametrization of time, Eqn.(\ref{gp8}) and finally, re-express these terms of matrix representation using the derivative of $\textbf{U}$. This gives
\bea 
 -{\cal E}=\frac{1}{2}\frac{1}{(N+1)^2 R^{2N}} (\sum ( U^{\prime}_i)^2) -\frac{\lambda^2}{8}R^{2N+2}-\frac{\mu}{R^{2N}} \label{gpce1}
 \eea
  Using this,  in Eqn.(\ref{gpPer5}) we get,
  
  \be
  \textbf{U}^{\prime \prime}-\frac{\lambda^2}{4}(2N+1)(N+1)^2 R^{4N}\textbf{U}+{\cal E}R^{2N-2}(2N)(N+1)^2 \textbf{U}=0\label{gpf} ,
  \ee
  where $ R^2= U^2_1+U^2_2$.  we re-write the above equation in $U_1$ and $U_2$ as
  \bea
  U^{\prime \prime}_{i}-\frac{\lambda^2}{4}(2N+1)(N+1)^2 (U^2_1+U^2_2)^{2N}U_{i}+{\cal E}(U^2_1+U^2_2)^{N-1}(2N)(N+1)^2 U_{i}=0, ~~i=1,2 \label{gpf1}
  \eea
  where a `prime' over $U$ stands for derivative with respect to the new time variable, $\tau$. These equations are obvious extensions of the Lagrangian
  \be
  {\cal L}=\frac{1}{2}m({U^{\prime}_1}^2+{U^{\prime}_2}^2)-{\cal E}m(N+1)^2 (U^2_1+U^2_2)^{N}+\frac{{\lambda}^2}{8}m(N+1)^2(U^2_1+U^2_2)^{2N+1} 
  \ee 
  
  Note that for $N=1$, these equations of motions lead to Eqn.(\ref{secho1}) and Eqn.(\ref{secho2}) i.e, equation of motions for perturbed harmonic oscillator problem and in addition $\lambda \rightarrow 0$, these describe motion of usual harmonic oscillator. Note that the $\lambda^2$ dependent terms are non-linear and these terms come due to the damping present in the original system.

  
\section{Kepler problem in 3-dimensions in presence of damping: Kustaanheimo-Stiefel transformation}  

The Kustaanheimo-Stiefel (KS) transformation, which is a three-dimensional extension of the Levi-Civita transformation, maps the  singular equations of motion of the three-dimensional Kepler problem to the linear and regular equations of a four-dimensional harmonic oscillator. Here we investigate the regularization of damped Kepler problem in 3-dimensions. We first give a brief summary of the definitions and identities involving quaternions, which are defined as
\bea
&{\underline U}=U_0+iU_1+jU_2+kU_3, ~
{\rm with}~i^2=j^2=k^2=-1,&\\
 &i\cdot j=-j\cdot i=k;~j\cdot k=-k\cdot j=i; ~k\cdot i=-i\cdot k=j&
\eea
We define the complex conjugate as
\be
{\bar {\underline U}}=U_0-iU_1-jU_2-kU_3
\ee
and a $*$-operation as
\be
{\underline U}^*=U_0+iU_1+jU_2-kU_3.
\ee
We note
\be
{\bar {\underline U}}{\underline U}= |{\underline U}|=\sum_{i=0}^3 U_{i}^2
={\bar {\underline U}^*}{\underline U}^*=|{\underline U}^*|^2 .
\ee
We also define a three vector 
\be
X=X_0+iX_1+jX_2\label{3vec}
\ee
which is given in terms of the quaternions using the KS transformation as
\be
X={\underline U}~{\underline U}^*.\label{kst}
\ee
Since ${\bar X}=X$, we find $(\bar{{\underline U}~{\underline U}^*)}=X$. One may write the above 
equation in the matrix form (with $X_3=0$) as
\bea
\left(\begin{array}{c}
X_0\\
X_1\\
X_2\\
X_3
\end{array}\right) =A(U)
\left(\begin{array}{c}
U_0\\
U_1\\
U_2\\
U_3
\end{array}\right),
\eea
where 
\bea
A(U)=\left(\begin{array}{cccc}
U_0&-U_1&-U_2&~U_3\\
U_1&~U_0&-U_3&-U_2\\
U_2&~U_3&~U_0&~U_1\\
U_3&-U_2&~U_1&-U_0
\end{array}\right).\label{Au}
\eea
This we can rewrite as follows,
\be
A(U)=\left(\mathcal{P}^{(0)}\textbf{U},\mathcal{P}^{(1)}\textbf{U},\mathcal{P}^{(2)}\textbf{U},\mathcal{P}^{(3)}\textbf{U} \right),\label{perks1}
\ee
where we have considered following set of permutations,
\bea
\mathcal{P}^{(0)}\textbf{U}=(U_0,U_1,U_2,U_3)^{T}, ~~~ \mathcal{P}^{(1)}\textbf{U} = (-U_1,U_0,U_3,-U_2)^{T}
 \nonumber \\
 \mathcal{P}^{(2)}\textbf{U}=(-U_2,-U_3,U_0,U_1)^{T}, ~~~ \mathcal{P}^{(3)}\textbf{U}=(U_3,-U_2,U_1,-U_0)^{T}.\label{perks2}
 \eea
 Here above used permutation matrices has some properties like,
 \be
 \mathcal{P}^{(i)}\mathcal{P}^{(j)} \neq  \mathcal{P}^{(j)}\mathcal{P}^{(i)} ~ and~~ (\mathcal{P}^{i})^{T}\mathcal{P}^{j}=\delta_{ij}.\label{perks3}
 \ee
We note that $X_3=0$, which naturally comes from the requirement ${\underline U}~{\underline U}^*=
({\underline U}~{\underline U}^*)^*$. We also note that $A(U)$ 
satisfy
\be
A(U)^TA(U)=r{\mathds I}.\label{Au1}
\ee
Defining $r=\sqrt{\sum_{i=1}^3 X_{i}^2}$, we find 
\be
r=\sqrt{|X|^2}=|{\underline U}|^2=|{\underline U}^*|^2.
\ee
Defining derivative as $\frac{\partial X}{\partial \tau}=X^\prime$, we find
\bea
{\underline U}~ d{\underline U}^*&=&d{\underline U}~{\underline U}^*\\
X^\prime&=&{\underline U}~{\underline U}^{\prime *}+ {\underline U^\prime}~{\underline U}^{ *}=2{\underline U}~{\underline U}^{\prime *\label{fd}}\\
X^{\prime\prime}&=&2{\underline U}^\prime~{\underline U}^{\prime *}\label{sd}\\
r^\prime&=&{\underline U}^\prime{\bar{\underline U}}+{\underline U}{\bar {\underline U}}^\prime
\label{rprime}
\eea
We also find 
\be
d\textbf{X}=2A(U)d\textbf{U}.\label{compat}
\ee
 Defining the momenta as
\be
\frac{dX_i}{dt}=P_i,~~\frac{dU_i}{d\tau}={\tilde p}_i, i=0,1,2,3.
\ee
shows that the compatibility with Eqn.(\ref{compat}) sets
\be
\frac{d\tau}{dt}=\frac{1}{4r}.\label{tre-para}
\ee
It can be verified easily that the anzats
\be
P=\frac{1}{2r}A(U){\tilde p}\label{Pptrans}
\ee
guarantee that the transformation $(X_i, P_i)\to (U_i, {\tilde p}_i)$ is a canonical transformation.

This is manifesting the generalization of Levi-Civita's conformal
squaring.
Hence the KS transformation maps ${\underline U} = (U_0, U_1, U_2, U_3) \in {\Bbb R}^4$ to 
$X = (X_0, X_1, X_2) \in {\Bbb R}^3$.
Since the mapping does not preserve the dimension, its inverse in the usual sense
does not exist.

\smallskip

Now, instead of considering the squaring of complex numbers as in the case of Levi-Civita regularization, we define mapping
in (\ref{kst}).  The image $ U \mapsto X := UU^{\ast}$ is the set of quaternions with vanishing $k$ components, which can be identified with ${\Bbb R}^3$.
By direct computation (\ref{kst}) yields
\be
X_0 = U_{0}^{2} - U_{1}^{2} - U_{2}^{2} + U_{3}^{2}, \,\,\,\, X_1 = 2(U_0U_1
 -U_2U_3), \,\,\,\, X_2 = 2(U_0U_2 + U_1U_3),\label{kst1}
\ee
follows directly from $X := UU^{\ast}$, thus $X = X^{\ast}$.
This is exactly the K-S transformation in its classical form.
Denoting the space coordinates by $(X_0, X_1, X_2)$ with
conjugate momenta $(P_0,P_1,P_2)$, the symplectic form becomes
$$
\omega = dX_0 \wedge dP_0 +  dX_1 \wedge dP_1 +  dX_2 \wedge dP_2 
$$
hence we can write $\left\lbrace X_l,P_n  \right\rbrace =\delta_{ln}$ for $l,n=0,1,2$. We must recall that 
$\omega$ generates the folded symplectic structure \cite{DKM}, if we choose $U_3 = 0$, then
$$
\omega = 2 ( U_0 dU_0 - U_1dU_1 - U_2dU_2) \wedge dP_0 + (U_0 dU_1 + U_1 dU_0 ) \wedge dP_1 + (U_0dU_2 + U_2dU_0) \wedge dP_2,
$$
which yields $ \omega \wedge \omega \wedge \omega = U_0(U_{0}^{2} - U_{1}^{2} - U_{2}^{2}) dU_0 \wedge dP_0 \wedge dU_1 \wedge dP_1 
 \wedge dU_2 \wedge dP_2$, hence this is a hyperbolic like $m$-folded symplectic structure. A folded symplectic structure is 
 a closed 2-form which is nondegenerate except on a hypersurface.

\smallskip

By direct computation one can check
\be
dX = d(UU^{\ast}) = dUU^{\ast} + U^{\ast}dU=
2dUU^{\ast} = 2U^{\ast}dU,\label{kst2} 
\ee
provided the `{\bf k}' component vanishes, which yields a bilinear relation \cite{StiSch,K-S,Waldvogel},
\be
U_0dU_3 - U_3dU_0 + U_2dU_1 - U_1dU_2 = 0. \label{bl}
\ee
This condition appears since KS transformation is a mapping from ${\Bbb R}^4$ to 
${\Bbb R}^3$, it therefore, leaves
one degree of freedom in the parametric space undetermined. By imposing this bilinear relation (\ref{bl}), the tangential map in (\ref{kst1}) change into a linear map and also this yields bilinear constraint \cite{StiSch,K-S,Waldvogel}, $U_3\tilde{p}_0 -U_2\tilde{p}_1 + U_1\tilde{p}_2 -U_0 \tilde{p}_3=0$, which makes $P_3=0$. 

Again by taking poisson bracket operation between coordinates, we get,
\be
\left\lbrace U_i,\tilde{p}_j  \right\rbrace =\delta_{ij} ~~ i,j=0,1,2,3. 
\ee 
 These relations implies that the mapping in (\ref{kst}), is canonical in nature.

\subsection{ Damped Kepler Problem}
 In this subsection, we study the regularization of collision orbits of a particle's motion in $\frac{1}{r}$ potential that is also exposed to a damping force in 3-dimensions, which follow from a time dependent Lagrangian.
We first use a time dependent point transformation to turn these equations into Euler-Lagrange equations derived from this time independent Lagrangian. This helps us to construct conserved energy, which then allows us pursuance of the mapping of dynamics on the constant energy surface.
After re-expressing these equations in terms of quaternions, we apply the K-S map to them. The equations are shown to represent a harmonic oscillator with inverted sextic potential and interactions.

We start with the Lagrangian  describing Damped Kepler problem in 3-dimensions
\be
L= e^{\lambda t}\left(\frac{m}{2}{\dot x}_{i}^2+\frac{k}{\tilde{r}}e^{-\frac{3\lambda t}{2}}\right),~i=0,1,2 \label{Lkep}
\ee
where $\tilde{r}=\sqrt{x_{i}^2}$ and ${\dot x}_i=\frac{d x_i}{dt}$. Euler-Lagrange equation following from this Lagrangian is
\be
m{\ddot x}_i+\lambda m {\dot x}_i+\frac{k}{\tilde{r}^3}e^{-\frac{3\lambda t}{2}}=0,~i=0,1,2 \label{eomk}
\ee
Applying the transformations
\be
x_i\to X_i=x_ie^{\frac{\lambda t}{2}},~i=0,1,2
\ee
we re-write the above Lagrangian in terms of new co-ordinates $X_i$ as
\be
{\cal L}=\frac{m}{2}\left ( {\dot X}_{i}^2+\frac{\lambda^2}{4}X_{i}^2-\lambda X_i {\dot X}_i\right ) +\frac{k}{r},\label{EqDKlag}
\ee
here $ r=\sqrt{X_{i}^2}$ and we have used summation convention. Note that the damping parameter $\lambda$ in Eqn(\ref{Lkep}),Eqn(\ref{eomk}) is now appearing a the co-efficient of $X_{i}^{2}$ and $X_{i}\dot{X_{i}}$ terms.

Equations of motion following from this Lagrangian are
\be
m{\ddot X}_i-\frac{m\lambda^2}{4}X_i+\frac{k}{r^3} X_i=0\label{eleqn},~i=0,1,2
\ee
We next derive the corresponding Hamiltonian corresponding to this Lagrangian. Calculating the conjugate momentum using
\be
P_i=\frac{\partial{\cal L}}{\partial{\dot X}_i}=m{\dot X}_i-\frac{m\lambda}{2}X_i,~i=0,1,2
\ee
and using this we obtain
\be
H=\frac{P_{i}^2}{2m}+\frac{\lambda}{2}X_iP_i-\frac{k}{r}\label{dkham},
\ee
(Here and below we have used summation convention)
which is a constant of motion. 
For later purposes, we express the total Hamiltonian in terms of the velocities as
\be
E=\frac{m}{2}{\dot X}_{i}^2-\frac{m\lambda^2}{8}X_{i}^2-\frac{k}{r}
\ee
and define
\be
-{\cal E}=-\frac{E}{m}=\frac{{\dot X}_{i}^2}{2}-\frac{\lambda^2}{8}X_{i}^2-\frac{\mu}{r}\label{com}
\ee
which is a constant of motion.

Next, we define a re-parametrization of time through the relation
\be
\frac{d}{d t}=\frac{1}{4r}\frac{d}{d\tau}\label{trepara}
\ee
and thus we find
\be
{\dot X}_i=\frac{1}{4r}X_{i}^\prime;~~{\ddot X}_i=\frac{1}{16r^2}X_{i}^{\prime\prime}-\frac{1}{16r^3}r^\prime X_i^\prime
\ee
Using the above re-parametrization, we re-express the equations of motion (\ref{eleqn}) as
\be
\frac{X^{\prime\prime}}{r^2}-\frac{1}{r^2} r^\prime X^\prime-4\lambda^2 X+\frac{16\mu}{r^3}X=0,\label{EQM}
\ee
where $\mu=\frac{k}{m}$. Here, we relabel $X_1\to X_0, X_2\to X_1$ and $X_3\to X_2$ for notational compatibility ( see eqn.(\ref{3vec})).

Now, using Eqns.({\ref{fd},\ref{sd},\ref{rprime}), we re-express eqn.(\ref{EQM}) as
\be
2r{\underline U}~{\underline U}^{\prime\prime*}-2|{\underline U}^\prime|^2~{\underline U}~{\underline U}^* +16\mu {\underline U}~{\underline U}^*
-4r^3\lambda^2 {\underline U}{\underline U}^*=0\label{reparaEQM}
\ee
Next, we apply re-parametrization to the  conserved quantity,  $-E$  in eqn.(\ref{com}) and re-express it as
\be
-{\cal E}=\frac{1}{32r^2}|X^\prime|^2-\frac{\lambda^2}{8} r^2-\frac{\mu}{r}
\ee
which in terms of quaternions become $-{\cal E}=\frac{1}{8r}|{\underline U}^{\prime *}|^2-\frac{\lambda^2}{8}r^2-\frac{\mu}{r}$ and we find
\be
-16{\cal E}r=2|{\underline U}^{\prime *}|^2-2\lambda^2r^3-16\mu.\label{hr}
\ee
Using this, we obtain from  eqn(\ref{reparaEQM}) 
\be
{\underline U}^{\prime\prime *} +\left(8{\cal E}-3\lambda^2r^2\right){\underline U}^*=0.\label{sho}
\ee
Since $r=|{\underline U}|^2$, we have $r^2= |{\underline U}|^4$ and using this, we re-express the above equation, explicitly, in components, as
\be
{U}_{i}^{\prime\prime } +\left(8{\cal E}-3\lambda^2 |{\underline U}|^4 \right){ U}_{i}=0\label{sho1}, ~i=0,1,2,3.
\ee
These equations are the Euler-Lagrange equations following from the Lagrangian
\be
{\cal L}_{SO}=\frac{1}{2}m(U_{i}^\prime)^2-4m{\cal E}U_{i}^2+\frac{m\lambda^2}{2}(U_{i}^2)^3\label{sextic}, ~i=0,1,2,3.
\ee

This Lagrangian, describes a harmonic oscillator with sextic potential and couplings in 4-dimensions. The Hamiltonian following from this is given by
\be
H=\sum_{i=0}^{3}\frac{\tilde{P_{i}}^2}{2m}+4m{\cal E}(\sum_{i=0}^{3}U_{i}^2)-\frac{m\lambda^2}{2}(\sum_{i=0}^{3}U_{i}^2)^3.\label{sexticH}
\ee
Note that the equations of perturbed Kepler problem in 3-dimensions given in Eqn.(\ref{eleqn}) is mapped under K-S transformation and re-parametrization of time variable, to equations of Harmonic oscillator with inverted sextic potential and couplings. Here too, we see that the regularized equation is non-linear and all the non-linear terms have $\lambda$ as the co-efficient, indicating that their origin is due to the damping in the original system.

\subsection{Homogeneous Hamiltonian}

 We now establish the equivalence between the damped Kepler problem in 3-dimensions to  harmonic oscillator with inverted sextic potential and interactions in 4-dimensions at the level of Hamiltonians. This is shown using homogeneous Hamiltonian formalism. 
 
 Homogeneous Hamiltonian formalism\cite{StiSch, Amer, gitmanbook,aaj} treats both space co-ordinates and time on an equal footing. This is achieved (i) by treating the time parameter $t$ as a function of another parameter, say $s$ ( then, $x(t)\equiv x(s)$), and (ii) by introducing a conjugate momentum for the time $t(s)$. This approach has been shown to be suitable to derive relativistic wave equations, using the usual replacement of momentum operators with the derivatives with respect to conjugate variables\cite{aaj}. This approach provide a natural way to study the quantization of systems with time dependent constraints\cite{gitmanbook}. This approach naturally fits into the canonical treatment of regularization transformations. In both Levi-Civita transformation and K-S transformation, time is treated as a function of new parameter. Using this framework, one can  easily show that the
 regularization transformations are canonical transformations\cite{StiSch}. Homogeneous formalism of non-relativistic H-atom was used to show its equivalence to harmonic oscillator in 4-dimensions\cite{grin}.
 
 We have seen that the damped Kepler problem in 3-dimensions is equivalent to the system described by the Lagrangian in Eqn.(\ref{EqDKlag}) and here we show that homogeneous 
 Hamiltonian corresponding to this model is equivalent to the Hamiltonian for 4-dimensional oscillator with inverted sextic potential and interactions. Note that the Lagrangian obtained 
 by time-dependent point transformation of damped Kepler problem in 3-dimensions given in Eqn.(\ref{EqDKlag}) is equivalent to
\be
{\cal L}=\frac{m}{2}\left ( {\dot X}_{i}^2+\frac{\lambda^2}{4}X_{i}^2 \right) +\frac{k}{r}\label{EqDKlag2}
\ee
as they differ only by a total derivative term and thus both lead to same equations of motion. The Hamiltonian derived from the above Lagrangian is
\be
{\cal H}=\frac{1}{2m}P_{i}^2-\frac{\lambda^2 m^2}{8}X_{i}^2-\frac{k}{r}.
\ee
Starting from this Hamiltonian, we derive the Hamiltonian describing harmonic oscillator with inverted sextic potential and interactions. For this, we first define the corresponding homogeneous Hamiltonian as
\be
H^H=4r\left({\cal H}+P_t\right)=\frac{2r}{m}P_{i}^2-\frac{r\lambda^2 m^2}{2}X_{i}^2-4{k}+4P_tr.
\ee
We now re-express this Hamiltonian in terms of ${\tilde p}_i$ and $U_i$, where we also use
$r=\sum_{i=0}^{3}U_{i}^2$ and $\sum_{i=0}^{2}X_{i}^2=(\sum_{i=0}^{3}U_{i}^2)^{2}$, to obtain
\be
{\bar H}^H=\sum_{i=0}^{3}\frac{{\tilde p}_{i}^2}{8m}+4E(\sum_{i=0}^{3}U_{i}^2)-\frac{\lambda^2m^2}{8}( \sum_{i=0}^{3}U_{i}^2)^{3}+p_s.
\ee
In the above, we have identified $P_t$ with $4E$ and $4k$ with $-p_s$ which is the conjugate of the new time parameter $\tau$. Note that the above homogeneous Hamiltonian $\bar{H}^{H}$, is defined in 4-dimension. The corresponding non-homogeneous Hamiltonian is
\be
{\bar H}= \sum_{i=0}^{3}\frac{{\tilde p}_{i}^2}{2m}+4E(\sum_{i=0}^{3}U_{i}^2)-\frac{\lambda^2m^2}{2}(\sum_{i=0}^{3} U_{i}^2)^{3}\label{sextich}
\ee
and it describes inverted sextic oscillator with couplings in 4-dimensions.We see that this Hamiltonian in Eqn.(\ref{sextich}) is exactly same as one obtained in Eqn.(\ref{sexticH}) with identification,
\be
 {\tilde p}_i=\tilde{P_i},~~ E=m{\cal E} 
\ee

Note that  $U_{i}^2$ dependent term in the Hamiltonian in Eqn.(\ref{sexticH}) and in Eqn.(\ref{sextich}), is in the form of harmonic oscillator potential with identification of relation between energy of perturbed Kepler  problem and strength (angular frequency) of harmonic oscillator i.e, $4E=\frac{1}{2}m\omega_{0}^2$. Here in Eqn.(\ref{sexticH}) and Eqn.(\ref{sextich}), damping parameter $\lambda$ appear as the co-efficient of
 inverted sextic potential and couplings.

\section{Conclusion}

In this paper, we have studied the regularization of central force systems with damping, both in 2 and 3 dimensions. The systems we have analyzed are (i) Kepler problem in 2-dimensions with friction included, (ii) particle moving under the influence of generic power law potential of the form $r^{-2N/N+1}$, $0\le 2N/N+1<2$ subjected also to a force linear in velocities, and (iii) damped Kepler problem in 3-dimensions. We have shown that the Levi-Civita transformation do regularize the equations of motion of first two systems while K-S transformations does  the same for third system.  In the first and third cases, the regularized equations describe harmonic oscillator
with inverted sextic potential and interactions, in 2 and 4 dimensions, respectively. In the second case, regularized equation is that of a harmonic oscillator with potential and interaction that depend on the actual value of $N$, $0\le N<2$. In all the three cases, we note that the non-linear terms in the regularized equations have $\lambda$ as the co-efficient where $\lambda$ is the damping parameter in the original models.

The explicit time-dependence of damped systems studied here is reflected in the fact that the corresponding Lagrangians/Hamiltonians also have explicit time dependence and thus breaks time translation invariance that the undamped systems enjoy. Thus, in these undamped systems,  we do not have conserved energy as expected for dissipative systems. Existence of conserved quantity is essential for implementing Levi-Civita and/or K-S transformations. To overcome this obstacle, we have mapped the time-dependent equations (equivalently, Lagrangians)  describing the damped systems to equivalent systems without explicit time dependence(see Eqns.
(\ref{lag1}, \ref{gp4}, \ref{EqDKlag}). These systems do have conserved energies and we use them in applying regularization transformations, in three steps as discussed in the introduction. The regularized equations obtained here are non-linear, unlike their undamped counter-parts. Our results reduced  to the well known, undamped cases in the limit $\lambda\to 0$, as expected.

We note that  in the case of generic potential $r^{-2N/N+1}, 0\le 2N/N+1<2$ with damping  in 2-dimensions, the permutation operators(see Eqn.(\ref{gpPer2})) were still commuting as in the undamped case. But in the case of 3-dimensions,
we note that the permutation operators(see Eqn.(\ref{perks3})) are not commuting. This feature  of permutation operators mutually commuting is not due to damping and even when there is no damping, these operators in the 3-dimensions would be non-commuting. Thus the identities satisfied by the transformation of co-ordinates in 2-dimensions shown in\cite{budd} will not be valid for 3-dimensional Kepler problem. Thus generalization of K-S transformation to the case generic potential $r^{-2N/(N+1)}$ as this was done in 2-dimensions fails.

One could actually construct conserved quantities from the time dependent Lagrangians/Hamiltonians of the damped systems considered here by adapting/generalising the method used in\cite{Lemos}) for deriving such a conserved quantity in 1-dimensions.  These conserved quantities will reduce to the familiar conserved energy when the damping parameter $\lambda$ vanishes. But using these conserved quantities in 
applying regularization transformations will lead to equations that depend on both new and old time parameters.\\

In an interesting paper, Andrade et al \cite{ADP-CV} studied Levi-Civita regularization problem of
the Kepler problem on surfaces of constant curvature.  
We would also like to study the regularization problem of the  Kepler equation with a drag force on 
surfaces of constant curvature., both positive
and negative, ${\Bbb S}^2$ and ${\Bbb H}^2$ respectively.

\section*{Acknowledgements}
SKP  thank UGC, India for support through JRF scheme(id.191620059604).
Work by the author PG was supported by the Khalifa University of Science and
Technology under grant number FSU-2021-014.

\bigskip

\renewcommand{\thesection}{Appendix : A}
\section{Motion in $r^2$ potential with damping in 2-dimensions:  Bohlin-Sudman Map}
\renewcommand{\thesection}{A}
In this appendix, we study the application of Bohlin-Sudman mapping to the equations of motion describing a damped harmonic oscillator. After mapping these equations to that of a shifted harmonic oscillator by a time dependent point transformation, we re-express the equations in terms of complex co-ordinates. Then by  applying a re-parametrization of time followed by a co-ordinate change, we map these equations describing a dynamics on a constant energy surface to that of a
particle moving in $\frac{1}{r}$ potential. 

We start from the equations of motions
\be
{ q}_{i}^{\prime\prime}+\lambda q_{i}^\prime+\Omega^2q_i,~i=1,2\label{dho-eq}
\ee
describing damped harmonic motion in 2-dimensions\footnote{These equations follow from the BCK Lagrangian $
L=\frac{1}{2m}\left({\dot q}_{i}^2-\Omega^2q_{i}^2\right)e^{\lambda \tau}, i=1,2$}. 
Here $q_{i}^\prime=\frac{dq_i}{d\tau}$ and $ q_{i}^{\prime\prime}=\frac{d^2q_i}{d\tau^2}$. We now apply the time-dependent co-ordinate transformation
\be
x_i=q_{i}e^{\frac{\lambda \tau}{2}}, i=1,2\label{map1}
\ee
and re-write the above equations of motion as
\bea
x_{1}^{\prime\prime}+{\tilde{\Omega}}^2x_1&=&0,\\
x_{2}^{\prime\prime}+{\tilde{\Omega}}^2x_2&=&0,\label{sho-eq}
\eea
where ${\tilde{\Omega}}^2=\Omega^2-\frac{\lambda^2}{4}$. These are Euler-Lagrange equations following from the Lagrangian
 \be
L=\frac{m}{2}\left( x_{1}^{\prime 2}+ x_{2}^{\prime 2}\right) -\frac{m}{2}{\tilde{\Omega}}^2(x_{1}^2+x_{2}^2)-\frac{m\lambda}{2}(x_1 x_{1}^\prime+x_2  x_{2}^\prime).\label{lag2}
\ee
We re-express these equations using  complex co-ordinate
\be
\omega=x_1+ix_2\label{comc}
\ee
as
\be
{\omega^{\prime\prime}}+{\tilde{\Omega}}^2\omega=0.\label{comeqn}
\ee
We now apply Bohlin-Sudman transformation
\be
\omega\to Z=\omega^2.\label{BStrans}
\ee
and also implement re-parametrization of time\footnote{ Bohlin-Sudman transformation has been applied to derive the mapping between 2-dimension harmonic oscillator to 2-dimension Kepler problem. Here, angular momentum is conserved in both systems and demanding these constants of motion are proportional to each other, results in the relation between time parameters of these two systems.  In the present case too, angular 
momentum is conserved for damped harmonic oscillator as well as (damped) Kepler problem in 2-dimension.} using
\be
{\bar Z}\frac{dZ}{dt}=\frac{\bar \omega}{2}\frac{d\omega}{d\tau}\label{repara}
\ee
Using Eqn.(\ref{BStrans}) in Eqn.(\ref{repara}), we get
\be
\frac{d}{dt}=\frac{1}{4{\bar {\omega}}{\omega}}\frac{d}{d\tau}
\ee
and using this we find
\bea
{\dot Z}&=&\frac{dZ}{dt}=\frac{1}{2{\bar {\omega}}}\frac{d\omega}{d\tau}\\
{\ddot Z}&=&\frac{d^2Z}{dt^2}=\frac{1}{8{\bar{\omega}\omega}}
\left[\frac{1}{\bar{\omega}}\frac{d^2\omega}{d\tau^2}-\frac{1}{{\bar{\omega}}^2}\left(\frac{d{\bar\omega}}{d\tau}\right)\left(\frac{d\omega}{d\tau}\right)\right]
\eea
From Eqn.(\ref{comeqn}), we have $\omega^{\prime\prime}=-{\tilde{\Omega}}^2\omega$ and using this, we re-write the second equation in the above as
\bea
{\ddot Z}&=&-\frac{1}{8{\bar{\omega}\omega}}
\left[\frac{1}{\bar{\omega}}{\tilde\Omega}^2\omega+\frac{1}{{\bar{\omega}}^2}\left(\frac{d{\bar\omega}}{d\tau}\right)\left(\frac{d\omega}{d\tau}\right)\right]\\
&=&-\frac{Z}{8|Z|^3}
\left[\left(\frac{d{\bar\omega}}{d\tau}\right)\left(\frac{d\omega}{d\tau}\right)+  {\tilde\Omega}^2{\bar\omega}\omega\right]\label{Kepler1}
\eea

 We now derive the conserved ``energy" associated with the Lagrangian in Eqn.(\ref{lag2}) and re-express the terms in the $[~]$ appearing in the above equation. To this end,
 we first obtain the conjugate momenta  corresponding to $x_i$ as
 \be
p_i=m x_{i}^\prime-\frac{m\lambda}{2} x_i, i=1,2
\ee
and construct the Hamiltonian in terms of velocities as
\bea
H&=&\frac{m}{2}\left( x_{1}^{\prime 2}+ x_{2}^{\prime 2}\right)+\frac{m{\tilde\Omega}^2}{2}\left(x_{1}^2+x_{2}^2\right)\\
&=&\frac{m}{2}\left[ {\bar{\omega}}^\prime \omega^\prime+{\tilde\Omega}^2{\bar\omega}{\omega}\right]\equiv E.
\eea
Since $E$ above is a constant, we use it to re-express Eqn.(\ref{Kepler1}) as
\be
{\ddot Z}=-\frac{E}{4m}\frac{Z}{|Z|^3}.\label{eQe}
\ee
We note that with the identification of conserved $E$ with $4k=m{\tilde\Omega}^2$, the strength of Kepler potential,
\be
\frac{E}{4}=k\label{id1},
\ee
the Eqn.(\ref{eQe}) is the Kepler's equation in 2-dimension, written in the complex co-ordinate $Z=X_{1}+iX_{2}
$,  

Thus we have mapped the equation of `damped' harmonic oscillator in 2-dimension,  to the equation of motion corresponding to the (undamped) Kepler problem in 2-dimension.

We now start with the expression for energy of 2-dim Kepler system, described in terms of ${\bar Z}$ and $Z$ and re-express it in terms of ${\bar\omega}$ and $\omega$, i.e.,
\bea
E_{Kepler}&=&\frac{m}{2}\frac{d{\bar Z}}{dt}\frac{dZ}{dt}-\frac{k}{|Z|}\\
&=&\frac{m}{8}\left[\frac{1}{{\bar\omega}\omega}\frac{d{\bar\omega}}{d\tau}\frac{d\omega}{d\tau}\right]-\frac{k}{{\bar\omega}\omega}
\eea 
Using above equation and Eqn.(\ref{id1}), we get 
\be
-E_{Kepler}=\frac{m}{8}\left(\Omega^2-\frac{\lambda^2}{4}\right).\label{EKepler}
\ee
Thus we find
\begin{enumerate}
\item The equations of motion of 2-dimensional damped Harmonic oscillator in Eqn.(\ref{dho-eq}) are first mapped to equations of a shifted harmonic oscillator given in Eqn.(\ref{sho-eq})
which are then mapped to that of (undamped) Kepler problem in 2-dimensions.
\item The strength of the Kepler potential is related to the conserved energy of the shifted harmonic oscillator.
\end{enumerate}


\end{document}